\documentclass[twocolappendix, numberedappendix]{emulateapj}
\usepackage{epstopdf}
\usepackage{textcase}
\usepackage{amsmath}

\begin{document}
\title{Solar Hard X-ray Source Sizes in a Beam-Heated and Ionised Chromosphere}

\author{Aidan M. O'Flannagain$^{1}$, John C. Brown$^{2}$, and Peter T. Gallagher$^{1}$}
\affil{$^{1}$Astrophysics Research Group, School of Physics, Trinity College Dublin, Dublin 2, Ireland}
\affil{$^{2}$Astronomy and Astrophysics Group, School of Physics and Astronomy, University of Glasgow, Glasgow G12 8QQ, Scotland, UK}

\shorttitle{HXR Sizes in an Ionised Chromosphere}
\shortauthors{O'Flannagain, Brown, Gallagher}

\begin{abstract}
 Solar flare hard X-rays (HXRs) are produced as bremsstrahlung when an accelerated population of electrons interacts with the dense chromospheric plasma. HXR observations presented by \citet{kon10} using the \textit{Ramaty High-Energy Solar Spectroscopic Imager} (\textit{RHESSI}) have shown that HXR source sizes are 3--6 times more extended in height than those predicted by the standard collisional thick target model (CTTM). Several possible explanations have been put forward including the multi-threaded nature of flare loops, pitch--angle scattering, and magnetic mirroring. However, the nonuniform ionisation (NUI) structure along the path of the electron beam has not been fully explored as a solution to this problem. Ionised plasma is known to be less effective at producing nonthermal bremsstrahlung HXRs when compared to neutral plasma. If the peak HXR emission was produced in a locally ionised region within the chromosphere, the intensity of emission will be preferentially reduced 
around this peak, resulting in a more extended source. Due to this effect, along with the associated density enhancement in the upper chromosphere, injection of a beam of electrons into a partially ionised plasma should result in a HXR source which is substantially more vertically extended relative to that for a neutral target. Here we present the results of a modification to the CTTM which takes into account both a localised form of chromospheric NUI and an increased target density. We find 50~keV HXR source widths, with and without the inclusion of a locally ionised region, of $\sim$3~Mm and $\sim$0.7~Mm, respectively. This helps to provide a theoretical solution to the currently open question of overly-extended HXR sources.
\end{abstract}


\section{Introduction}

Solar eruptive events are the largest explosions in the solar system, releasing energy in the form of radiation and ejected matter on the order of $10^{32}$~ergs ($10^{25}$ Joules) in a matter of minutes \citep[e.g,][]{ems04, ems05}. Part of this emission takes the form of hard X-rays (HXRs). This high-energy emission is known to be produced by accelerated particles interacting with the corona and chromosphere, but the dependence of the spatial structure of this emission on the dynamics of the target plasma is still poorly understood. The Ramaty High Energy Solar Spectroscopic Imager \citep[RHESSI;][]{lin02} has helped to provide insight into this area since its launch in 2002. A number of studies have been performed on RHESSI source positions which tell us where in the dense chromosphere accelerated electrons lose their energy \citep[e.g.][]{asc02, bat12, ofl13}. While a good understanding has resulted in terms of the HXR source centroid, the vertical extent is still observationally found to be 
substantially underestimated by the Collisional Thick Target Model \citep[CTTM;][]{bro71, hud72}. Indeed, a detailed RHESSI study performed by \citet{kon10} aimed to measure the sub-arcsecond spatial properties of one particular solar flare that occured on 6 January 2004. In that work, the visibility forward fit (VIS FWDFIT) method was used, which by fitting model visibilities of pre--defined Gaussian sources to observed RHESSI visibilities, can produce moments of the HXR distribution with a greater accuracy than RHESSI's finest grid pitch \citep{sch02}. With this method, it was shown that observed vertical source extents were in the range of 3--6~arcsec, depending on energy. This was a consistent factor of 3--6 times more extended than their simulated counterparts \citep[see also][]{den09}. This presents a clear disconnect between observation and the most commonly--used model for solar flares.

The standard model of solar flares describes an energy release in the corona, believed to be magnetic reconnection \citep{swe58, pet64}. Particle acceleration results from either the strong electric fields associated with the changing magnetic field resulting from reconnection, or a secondary mechanism such as shocks or turbulence formed around the diffusion region (see \citet{zha11} for a recent review). These accelerated particles travel down the reconnected magnetic field lines into the dense chromosphere, where they lose their energy to Coulomb collisions and emit bremsstrahlung \citep[CSHKP model;][]{car64,stu66,hir74,kop76}. The CTTM describes the relationship between the HXR spectrum produced and the injected spectrum of electrons. It has also been used to make predictions on the nonthermal HXR footpoint vertical intensity distribution \citep{bro75, bro02}, and so can produce estimations of the vertical extent of chromospheric HXR sources for a given density structure along the flare 
loop. In order to reproduce sources which are 3--6 times larger than the standard CTTM predicts, as are observed by RHESSI, one could alter the density structure until the observed extent is reached. However, \citet{bat12} have shown that that this would require a strong enhancement of coronal density. This was shown to result in electrons of energy as high as $\sim$20~keV being stopped in the corona, failing to produce chromospheric footpoints. Additionally, the coronal densities required for a substantial increase in source size are larger than $\sim$$10^{11}$ $\textrm{cm}^{-3}$, which were noted to be non-typically high.

A further category of density models, which are based on simulating the response of a quiet-Sun plasma to an injection of nonthermal electrons, may provide a solution to this problem. Theoretical work invoking full radiative and hydrodynamic models shows that upon heating by an electron beam, a portion of the chromosphere rapidly becomes ionised \citep{fis85, abb99, all05}. It has been previously shown that variation in the ionisation fraction along the path of the beam, or non-uniform ionisation (NUI) should produce a break in the HXR spectrum, as ionised plasma is a less efficient bremsstrahlung target \citep{bro72, su09}. It has been demonstrated that a rise or fall in ion fraction across the transition region will cause a variation in the vertical structure of HXR emission \citep{ems78}. However, the effect of a local variation of ion fraction within the chromospheric target has not been explored.

In this work, we demonstrate how this reduction of HXR production efficiency in a locally ionised chromosphere, together with the increased density in the upper chromosphere can result in a substantial increase in the vertical extent of the emitted source size. We discuss this increase in the context of RHESSI's finest spatial observations. We also briefly demonstrate the effect this local ionisation has on the RHESSI spectrum, as this may contribute to the break normally attributed to the variation in ionisation profile encountered at the transition region.

\section{Method}

In this Section, we outline the details of expressions for HXR distributions in height (profiles; Section 2.1) and energy (spectra; Section 2.2) produced by a beam of nonthermal electrons injected into a plasma of freely-varying ionisation fraction and temperature. For the injected distribution $f_{0}$ of electron energies $E_{0}$, we will take the standard power-law form without a low-energy cutoff:
\begin{equation}
f_{0}(E_{0}) = (\delta - 1)\;\frac{f_{1}}{E_{1}}\;  \left(\frac{E_{0}}{E_{1}}\right)^{-\delta} 
\end{equation}
where $f_{1}$ and $E_{1}$ are a reference flux and energy, and $\delta$ is the spectral index. For the remainder of the paper, we will only consider HXR photons with energies greater than 20~keV, and will assume a low--energy cutoff below this value. As a low--energy cutoff will have no effect on emission with a greater photon energy, it is satisfactory to exclude one from our injected beam model.

\subsection{Target Density Model}

In this work, a modified form of the standard hydrostatic equilibrium (HSE) density structure is used. The standard form describes an exponential drop in density with height, z, above the photosphere: $n(z) = n_{ph} exp(-z/H)$, where $n_{ph}$ is the density at the top of the photosphere, and $H$ is the scale height. The scale height in this case assumes a constant temperature $T$ and ionisation fraction $X$ and has the form $H = k_{B}T/[(X+1)m_{p}g]$, where $g$ is the acceleration due to gravity at the solar surface. However, to take into account an arbitrary variation in temperature and ionisation fraction, we adopt a heuristic form for the exponent change such that
\begin{equation}
 n(z) = n_{ph} exp\left[-\frac{m_{p}g}{k_{B}}\int_{0}^{z}\frac{dz}{(X(z)+1)T(z)}\right]
\end{equation}
where $k_{B}$ and $m_{p}$ are Boltzmann's constant and the proton mass, respectively. In this case we have made the assumption that the dominant term in the expansion of $d(n(X+1)T)/dn$ is $(X+1)T$. This assumption may not hold in all cases for the solar atmosphere, but for the purposes of producing a density enhancement, this assumption is found to be satisfactory at least as an empirical means of exploring the NUI effect. This expression for density takes into account the rapid variation in scale height both at the transition region and as a result of localised beam heating and ionisation.

\subsection{HXR Height Profile}
The distribution of HXR emission along a flare loop leg in the chromosphere can be determined by taking into account collisional energy loss of an electron of initial energy $E_{0}$ to an energy $E$. The electron loses energy as it travels to a column depth $N(z)$, which is the integral of target density, $n(z)$ along the path of the electron: $N(z) = -\int^{\infty}_{z} n(z) dz$. This energy loss is given by $E_{0}^{2} - E^{2} = 2KN$, where $K = 2\pi e^{4} \Lambda$, and $\Lambda$ is the Coulomb logarithm, which accounts for the range of collision impact factors that result in significant energy loss, and so depends on the ionisation state of the plasma \citep{bro72}.

To arrive at a HXR spectrum of photons of energy $\epsilon$, this collisionally modified electron distribution, $f(E)$, is multiplied by Kramers bremsstrahlung cross-section $\sigma (\epsilon, E) = \sigma_{0} / (\epsilon E)  $\citep{kra23}. \citet{bro02} derived the distribution of nonthermal photon flux with height to be
\begin{align}
 \frac{dI}{dz} = \frac{Af_{1}\sigma_{0}}{8\pi r^{2}E_{1}}\; &(\delta - 1)\;\frac{1}{\epsilon}\;n(z) \; \left(\frac{E_{1}^{2}}{2KN(z)}\right)^{\delta/2}\;\nonumber\\ 
   &\times \; B\left(\frac{1}{1+u(z)},\frac{\delta}{2},\frac{1}{2}\right)
   \label{dIdz}
\end{align}
where $A$ is the cross-sectional area of the loop, $r$ is the distance to the observer, $u(z) = \epsilon^{2}/2KN(z)$ and B(...) is the Incomplete Beta Function
\begin{equation}
B\left(\frac{1}{1+u},\frac{\delta}{2},\frac{1}{2}\right) = \int^{1/(1+u)}_{0}{x^{\delta/2-1}\:(1-x)^{-1/2}\:dx}
\end{equation}


In order to account for variations in ionisation fraction $X$, we follow \citet{bro98} in defining an effective column depth $M$ such that our energy loss is
\begin{equation}
 E_{0}^{2} - E^{2} = 2 K' M
\end{equation}
\begin{equation}
 M = \int_{0}^{N} (\lambda + X(N')) dN'
\end{equation}
where $K' = 2\pi e^{4} \Lambda$, $\Lambda = \Lambda_{ee} - \Lambda_{eH}$, and $\lambda = \Lambda_{eH}/\Lambda \approx 0.55$ \citep{bro73}. By replacing the $2KN$ factors in Equation \ref{dIdz} with $2K'M$, we can produce a model HXR height profile which takes into account variation in ionisation fraction both at the transition region and within the target.

\subsection{HXR Spectrum}

The predicted HXR spectrum resulting from an electron beam passing through plasma which contains a step-function in ionisation fraction at the transition region has been given in a number of works \citep{bro73, bro98, su11}. This work was expanded upon to account for a region of linear variation from a fully ionised to a fully neutral target by \citet{su09}. In Appendix A we present a further expansion which allows for any possible variation in ionisation fraction.

\begin{figure}
 \includegraphics[scale = 0.45]{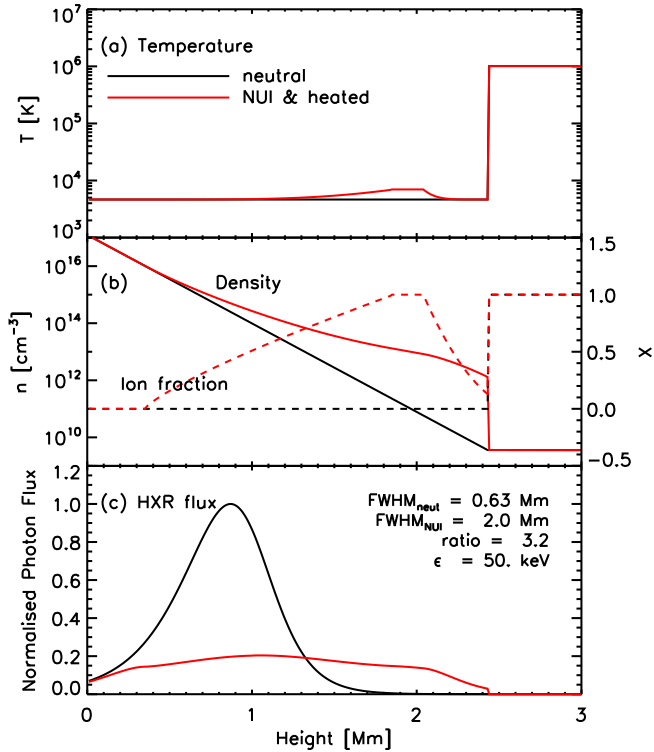}
 \caption{(a) Target ambient plasma temperature variation with height. The quiet-Sun case is shown in black, and the flaring target case is shown in red. In the latter case, a localised increase in temperature, peaking around 2~Mm, is present due to beam heating. The sudden rise in temperature around 2.5~Mm constitues the transition region. (b) Target plasma density (solid lines) and ionisation fraction (dashed line) distributions with height. In the flaring scenario, the chromosphere has been ionised around the region of peak beam energy deposition, which along with the rise in temperature causes an increase in scale height. This results in an enhancement of density in the upper chromosphere. (c) HXR flux distribution versus height for photons of 50~keV. The black and red flux profiles correspond to the neutral and NUI target cases, respectively, and both profiles are normalised to peak emission in the neutral case. Also shown are the full-width-half-max (FWHM) of both profiles, and their ratio.}
 \label{fig_target}
\end{figure}

\section{Results}

\subsection{HXR Height Profile}

The parameters of our target plasma, as well as a sample HXR height profile for photons of 50~keV are shown in Figure~\ref{fig_target}. Black lines in all cases describe a fully ionised corona and fully neutral chromosphere. Red lines show the atmospheric parameters and HXR fluxes when local heating and ionisation are taken into account using the model presented here. As shown in Figure~\ref{fig_target} (a) and (b), we have approximated a localised beam energy deposition with a rise in both temperature and ionisation fraction, which produce a density which is enhanced in the upper chromosphere, and decreases more gradually with height. As a result of the combination of this density enhancement and the reduction of HXR emission around the chromospheric ionised region, we see a strong increase in vertical extent of the 50~keV emission as shown in Figure~\ref{fig_target} (c).

\begin{figure}
 \includegraphics[scale = 0.45]{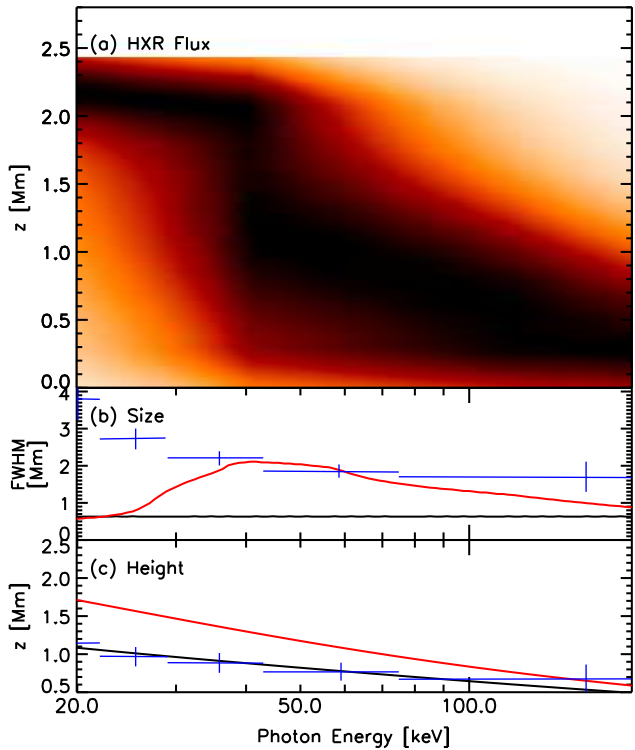}
  \caption{(a) Distribution of HXRs in height and energy, in the case where nonuniform ionisation is taken into account. Emission is normalised to the maximum flux in its energy band. 
  (b) Vertical size and (c) centroid height of the model sources versus energy, for the nonuniform (red) and fully neutral (black) case. Overplotted in blue are the observational values recorded in \citet{kon10}, as shown in Figure 6 of that paper.}
 \label{fig_result}
\end{figure}

Vertical sizes of the model sources are determined by taking the FWHM of emission.
As shown in Figure~\ref{fig_target} (c), the inclusion of NUI and heating effects increases the extent of our source from 0.63~Mm to 2.0~Mm, an increase by a factor of 3.2 for a 50~keV source. This effect can be seen for energies between 20~keV and 200~keV upon examination of Figure~\ref{fig_result}.

A complete set of resulting HXR height profiles are given in Figure~\ref{fig_result} (a), in the form of an image of HXR flux against both height and energy. That is, a vertical slice of Figure~\ref{fig_result} (a) is a height profile as presented in Figure~\ref{fig_target} (c).
 In Figure~\ref{fig_target} (b) and (c), the vertical size and centroid height of model HXR source are shown, respectively. Again, the red lines correspond to the NUI case and solid to the neutral chromosphere case. Also shown in blue are the observed values for size and height as measured using RHESSI, presented in \citet{kon10}, the work which initially highlighted the disparity between observed and predicted source vertical extents.

It is immediately apparent that, above $\sim$30~keV, a substantial increase in vertical extent results from the inclusion of NUI in the target plasma. The ratio of source FWHM in a NUI target to that of a neutral one reaches a peak of a factor of 3.3 at 40~keV, and slowly falls with increasing energy to a value of $\sim$2.1 at 100~keV, and $\sim$1.4 at 200~keV. For the energy range of $\sim$30--70~keV, the NUI-adjusted vertical extent matches those measured by RHESSI. Outside of this range, source sizes gradually return to that of the neutral case. This may be due to the fact that electrons emitting photons outside this energy range are penetrating to depths above and below the location of the peak in ionisation, and therefore result in standard neutral target emission. The source height on the other hand is much less strongly affected within the same energy range. As shown in Figure \ref{fig_result} (c), the height of the NUI source centroid is a factor of 1.6 and 1.5 higher than its neutral counterpart at 
20~keV and 40~keV, and this disparity continues to diminish with increasing energy.

\subsection{Instrumental Effects}

A further important contribution to HXR source size which requires discussion is the effect of RHESSI's imaging response. Due to the finite spacing of RHESSI's finest grids, it is expected to have a lower resolution limit, which may result in an increase in the apparent size of small sources. This point--spreading effect has previously been shown to be insufficient to completely account for the difference between predicted and observed source sizes \citep{bat12}. However, it is useful to determine if RHESSI would be able to detect the difference between the neutral and NUI HXR sources presented here.

Model two-dimensional HXR sources were produced by taking the matrix product of the model flare emission profiles and a normalised Gaussian with a FWHM of 3~arcsec. The resulting maps were then used as input to RHESSI simulation software developed by R. Schwartz \citep[for example see][]{sui02}. This software produces a RHESSI calibrated eventlist from an input simulated map \citep{sch02}. From the eventlist, any of the available standard image reconstruction methods can then be used to produce an image. For our purposes, we use the VIS FWDFIT method \citep{sch07}. VIS FWDFIT uses a characteristic shape for an X-ray source such as a circular or elliptical Gaussian, and fits the resulting simulated visibilities to those observed by RHESSI. The resulting Gaussian parameters, along with their standard deviations are returned as an output, and can be conveniently used to directly measure properties such as position and extent of the HXR source.

In order to accurately compare with the observations made by \citet{kon10}, we used the same set of detectors in the reproduction of RHESSI images from the synthetic input maps. These were detectors 2--7, as in that observation, detector 1 showed no significant signal, and detectors 8 and 9 were deemed too coarse for the event. In addition, the input total counts parameter was chosen in order to match that of the flare which occured on 6 January 2004, in the timeframe used to produce images in that study. This was estimated to be 7 $\times$ $10^{3}$ counts detector$^{-1}$, as that was roughly the number of counts collected in each of the energy bands used in \citet{kon10}. This selection was made in an effort to demonstrate that any distinction between sources emitted by a neutral and NUI target would be measurable by RHESSI in a real flare.

\begin{figure}
 \includegraphics[scale = 0.45]{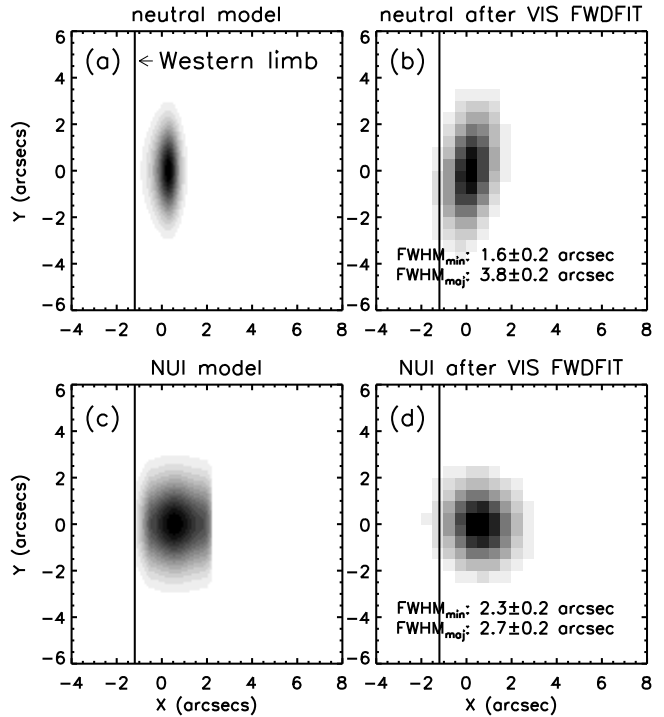}
 \caption{Demonstration of the effect of RHESSI's imaging response to HXR sources produced by neutral and NUI target chromospheres. (a/c): A model HXR input map produced by taking the matrix product of a 3~arcsec FWHM Gaussian with the neutral/NUI HXR profile of 50~keV emission as shown in Figure \ref{fig_target} (c). The vertical line represents the western solar limb. The sharp dropoff in flux at the right of the source in (c) is due to the sharply--defined transition region, shown at $\sim$2.4~Mm in Figure \ref{fig_target} (c). (b/d): The map produced by running the VIS FWDFIT routine on the calibrated eventlist resulting from the map shown in (a). This serves as an approximation of a RHESSI image given a HXR source produced in a neutral/NUI chromosphere.}
 \label{model_maps}
\end{figure}

\begin{figure}
 \includegraphics[scale = 0.45]{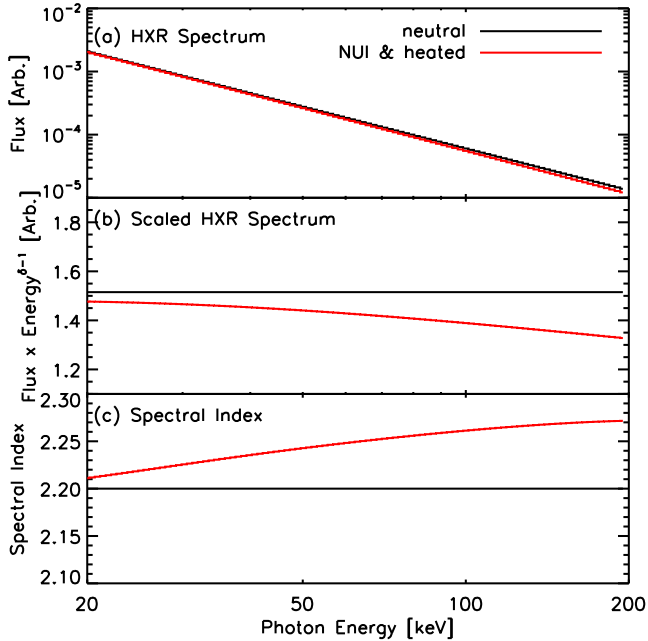}
  \caption{(a) Model HXR spectra resulting from the interaction of a power-law beam of electrons with a neutral (black) and nonuniformly ionised (red) target. (b) Same data as (a), but flux is multiplied by $\epsilon ^{\delta-1}$ in order to more clearly show the distinction between the two cases. (c) The spectral index of the HXR spectra shown in (a), calculated by using the derivative of log(Flux) with respect to log(Photon Energy).}
 \label{fig_spec}
\end{figure}

The result of running VIS FWDFIT on model 50~keV HXR sources emitted by a neutral and NUI target are shown in Figure~\ref{model_maps}. Input maps are shown in panels (a) and (c), and the maps resulting from running VIS FWDFIT on the calibrated eventlist produced by these input maps are shown in (b) and (d). Given that this simulated observation occurs on the west limb of the Sun, the vertical line represents the rough location of the base of the chromosphere, based on the model given in this paper. As shown, both sources are radially broadened to a degree by the simulated RHESSI imaging process. There still remains a clear distinction in source size between the neutral sources and those produced in a density--enhanced and NUI region for the realistic detector choice and countrate used. The FWHM of the minor axes of the neutral and NUI sources after instrumental effects are 1.6 and 2.3~arcsec, with standard deviations of 0.2~arcsec in both cases. It could be inferred that sources of 
FWHM$<$1.6~arcsec are extended up to this limit, but sources that are already larger are unaffected. This highlights the requirement for a physical mechanism to increase the vertical HXR extent in order to match observations. It is noted here that these measurements should be treated as lower limits only, due to the fact that the calibrated eventlists are produced using the same grid parameters that are then used to reproduce images \citep[see for example,][]{bat11b}.

\subsection{HXR Spectrum}

The HXR spectrum resulting from interaction between our power-law injected spectrum of electrons and the NUI plasma shown in Figure~\ref{fig_target} is shown in Figure~\ref{fig_spec} (a). As no low-energy cutoff has been included in this model, the resulting HXR photon spectrum appears to be a straightforward power-law spectrum with a spectral index that is constant with energy. However, the disparity is made clear in the rescaled spectra in \ref{fig_spec} (b). As shown upon examination of Figure~\ref{fig_spec} (c), there is some variation in photon spectral index. As energy rises to 200~keV, the index of the spectrum rises from $\sim$2.20 to $\sim$2.27, a rise of $\sim$0.07. This variation is qualitatively similar to that presented by \citet{su09}, but in terms of the magnitude of the variation in index, is far smaller than the observed average rise of $\sim$0.3--1.4. This may suggest that local ionisation within the chromosphere contributes some small amount to the known break in RHESSI spectra, 
but the major cause of the break which can be attributed to NUI would still be the variation in ionisation fraction across the transition region.

It should however be noted that there are methods of producing a break in the HXR spectrum which serve as an alternative to NUI. For example, incorrect background subtraction of the solar spectrum could result in a photon spectrum which does not relate to the accelerated electron spectrum, or pulse pile-up may not be corrected for adequately \citep{smi02}. Beyond instrumental effects, HXR albedo \citep{kon10alb, kon06}, return currents \citep{kni77, hol12}, or the presence of a low-energy cutoff in the injected spectrum \citep{gan02, sui07} may all also contribute to a break in the usual power-law HXR spectrum.

\section{Conclusion}

Here we have shown that the predicted local ionisation and heating of chromospheric plasma on the onset of a flare can produce an increase in the vertical HXR source extent by a factor of up to $\sim$3 for commonly-imaged energies. This increase results from a combination of localised ionisation and a density enhancement in the upper chromosphere, but does not require any increase in coronal density, and therefore results in no extreme upward shift of source centroids. This helps to explain the observation of RHESSI source sizes that appear too large when compared to predictions based on simple interpretations of the CTTM. In particular, the red curve in Figure~\ref{fig_result} (b) can be directly compared with the measured values from \citet{kon10}, shown in blue. In both the model and observed cases, the HXR source size decreases with increasing energy, however model source sizes were shown to be a factor of 3--6 smaller in vertical extent than observations. By taking into acount local NUI in addition to a 
chromospheric density enhancement, we have been able to account for most of this discrepancy which, while resulting in some increase in source height, does not encounter the problem of dramatically relocating HXR sources to the corona.

Previously, the disagreement between observed and modelled source sizes has been addressed by a number of mechanisms. \citet{kon10} point out that treating the flare loop as a single monolithic loop is an oversimplification, and perhaps strands of different density structure could contribute to an apparent vertical lengthening of the source. Various physical processes expected to take place during flares, such as magnetic mirroring and pitch-angle scattering, have also been put forward as other contributing factors. However, Fokker-Planck modelling of the nonthermal electron distribution in a flare has demonstrated that these processes would have only minor effects on source size \citep{bat12}. The resulting conclusion has been that a threaded loop structure remains the best explanation. While these processes are certainly still expected to take place, they may no longer be required in order to explain the observed discrepancy in vertical HXR source sizes.

It is important to clarify here the mechanism by which the NUI component of this model, aside from the density enhancement, produces a more vertically extended source. In addition to causing an enhancement in the wings of a HXR source profile such as that shown in Figure~\ref{fig_target} (c), the ionisation of the plasma also has the effect of reducing emission near the peak. By reducing the efficiency of emission primarily near this peak, the overall source becomes broader. Because, for lower photon energies, peak emission and peak ionisation both occur at the same position along the loop, it is expected that the reduction will always occur at the location of brightest HXR flux, and therefore always produce some extension. This suggests that the effects described here should be common, and possibly applicable to all flares.

A noteworthy limitation of this model is that hydrostatic equilibrium is assumed immediately following the injection of energy by a beam of relativistic electrons. In reality, the intense localised heating should result in rapid expansion of local chromospheric plasma, and a mostly upward-directed pressure wave \citep{fis85, abb99, all05}. In order to fully assess the effect of the dynamic variation of temperature, density, and ionisation fraction on HXR emission, a full radiative and hydrostatic model should be implemented which, at each time interval, produces a HXR height profile and spectrum.

\begin{appendix}

\section{Generalised HXR NUI Spectrum}
 Following the work listed above we have the general form for distribution of HXR flux with energy, assuming again that the Kramers bremsstrahlung cross-section is adequate:
\begin{equation}
 J(\epsilon) = \frac{1}{4\pi r^{2}} \frac{\sigma_{0}}{\epsilon K'} \int^{\infty}_{\epsilon} \int^{\infty}_{E} \frac{f_{0}(E_{0})}{\lambda + X(M)} dE_{0} dE
 \label{spec_basic}
\end{equation}
where all symbols have their previous meaning. We introduce an ionisation fraction $X(M)$ which will vary arbitrarily over $L$ linear steps:
\begin{equation}
X(M) =
\left\{ \begin{array}{ll}
         X_{0} & \mbox{: $0 \leq M < M_{1}$}\\
         X_{1} & \mbox{: $M_{1} \leq M < M_{2}$}\\
         \vdots\\
         X_{L} & \mbox{: $M_{L} \leq M < M_{L+1}$}
         \end{array} \right.
\end{equation}
As the limit of the inner integral of Equation \ref{spec_basic} is a function of $X(M)$, we take into account variation of $X(M)$ over $L$ linear steps by breaking up the integral into $L$ terms.
\begin{multline}
 J(\epsilon) = \frac{1}{4\pi r^{2}} \frac{\sigma_{0}}{\epsilon K'} \int^{\infty}_{\epsilon}
 \left[\int^{\sqrt{E_{0}^{2} + 2K'M_{1}}}_{E} \frac{f_{0}(E_{0}')}{\lambda + X_{0}} dE_{0}' \right.\\ 
 +\left.\int^{\sqrt{E_{0}^{2} + 2K'M_{2}}}_{\sqrt{E_{0}^{2} + 2K'M_{1}}} \frac{f_{0}(E_{0}')}{\lambda + X_{1}} dE_{0}' \right. \\
  \cdots + \left.\int^{\infty}_{\sqrt{E_{0}^{2} + 2K'M_{L}}} \frac{f_{0}(E_{0}')}{\lambda + X_{L}} dE_{0}'  \right]dE
 \label{spec_break}
\end{multline}
We have essentially arrived at Equation 4 of \citet{su09}, however in this case we have accounted for an arbitrary number of linear steps in $X(M)$ rather than one. This leads us to the following general expression of the HXR spectrum, following the substitution of our power-law injection spectrum, given at the beginning of this section.

\begin{multline}
 J(\epsilon) = \frac{f_{1}}{E_{1}^{1-\delta}}\frac{1}{4\pi r^{2}} \frac{\sigma_{0}}{\epsilon K'}
 \left[ \frac{E^{2-\delta}}{(2-\delta)(\lambda + X_{0})}\right.\\ 
 +\left. \sum\limits_{l = 1}^{L}\left( \frac{1}{2(\lambda + X_{l-1})}\left(2K'M_{l}\right)^{-\delta + 1/2}B\left(\frac{1}{1+\frac{\epsilon^{2}}{2K'M_{l}}},\frac{\delta}{2},\frac{1}{2}\right)\right.\right.\\
 +\left.\left. \frac{1}{2(\lambda + X_{l})}\left(2K'M_{l}\right)^{-\delta + 1/2}B\left(\frac{1}{1+\frac{\epsilon^{2}}{2K'M_{l}}},\frac{\delta}{2},\frac{1}{2}\right)\right)\right]
 \label{spec_final}
\end{multline}

This expression can now be used to produce a HXR spectrum for any given variation in ionisation fraction along the path of the nonthermal beam, assuming a finite number of linear steps is an acceptable approximation of this variation. We are now able to produce both model HXR height profiles and spectra for any input ionisation fraction $X(M)$.
\end{appendix}

\begin{acknowledgements}
  We thank M. Battaglia for her valuable input on RHESSI's imaging response. This work has been supported by a Government of Ireland Studentship (AMO'F) from the Irish Research Council (IRC) and a Visiting Lectureship by TCD (JCB).
\end{acknowledgements}

\bibliographystyle{apj}
\bibliography{paper2.bib}

\end{document}